# Impurity effects in coupled-ladder BiCu$_2$PO$_6$: NMR and QMC

L. K. Alexander, [1] J. Bobroff, [1*] A. V. Mahajan, [2] B. Koteswararao, [2] N. Laflorencie [1], F. Alet [3]

[1]*Laboratoire de Physique des Solides, Univ. Paris-Sud, UMR8502, CNRS, F-91405 Orsay Cedex, France*
[2]*Department of Physics, Indian Institute of Technology Bombay, Mumbai 400076, India*
[3] *Laboratoire de Physique Théorique, IRSAMC, Université de Toulouse, CNRS, F-31062 Toulouse, France*

23 December 2009

We present a $^{31}$P NMR study of the coupled spin ½ ladder compound BiCu$_2$PO$_6$. In the pure material, intrinsic susceptibility and dynamics show a spin gap of about $\Delta \approx 35-40K$. Substitution of non magnetic Zn or magnetic Ni impurity at Cu site induces a staggered magnetization which results in a broadening of the $^{31}$P NMR line, while susceptibility far from the defects is unaffected. The effect of Ni on the NMR line broadening is twice that of Zn, which is consistent with Quantum Monte Carlo (QMC) calculations assuming that Ni couples ferromagnetically to its adjacent Cu. The induced moment follows a *1/T* temperature dependence due to the Curie-like development of the moment amplitude while its extension saturates and does not depend on impurity content or nature. This allow us to verify the generically expected scenario for impurity doping and to extend it to magnetic impurity case: in an antiferromagnetically correlated low dimensional spin system with antiferromagnetic correlations, any type of impurity induces a staggered moment at low temperature, whose extension is not linked to the impurity nature but to the intrinsic physics at play in the undoped pure system, from 1D to 2D systems.

## I. INTRODUCTION

Studying the effects of impurity substitutions in low-dimensional magnetic systems is a productive method to reveal the magnetic properties of the pristine, undoped materials as well as to probe the unusual and novel short-range magnetic disturbance that a dopant/impurity might cause around it. In particular, nuclear magnetic resonance (NMR) experiments have been successful in unravelling the details of the impurity-induced short-range and long-range effects via a measurements of the NMR lineshapes and their dependencies on temperature.[1] A direct comparison with theoretical predictions of the ground-state properties is then possible. The prominent systems which have been investigated are the *S* = ½ (gapless) Heisenberg Antiferromagnetic (HAF) chain Sr$_2$CuO$_3$ [2], the *S* = ½ (spin-gapped) spin-Peierls chain CuGeO$_3$ [3], the *S* = 1 (spin-gapped) HAF Haldane chain Y$_2$BaNiO$_5$ [4], the *S* = ½ (spin-gapped) 2-leg ladder SrCu$_2$O$_3$ [5,6] etc. From the theory side, impurity-induced effects in low-D antiferromagnets has been studied quite extensively as well [7-15].

Continuing in this endeavor, we have recently reported[16] the occurrence of a spin-gap behaviour in the *2*-leg ladder compound BiCu$_2$PO$_6$. This was confirmed by recent neutron experiments [17]. The most studied spin-ladder compound (SrCu$_2$O$_3$) shows a very small coupling between ladders accompanied by a very large coupling *J* = *2000 K* between the Cu spins in the ladder, i.e. it consists of almost isolated ladders. BiCu$_2$PO$_6$ on the contrary shows a smaller coupling in the ladder *J = 100 K* but comparatively larger coupling between the ladders[16,17]. This allows to probe the intermediate situation between isolated ladders and 2D-Heisenberg AF planes which should help understand how peculiar the ladder geometry is and what the properties of such coupled ladders are. In this spirit, we recently reported μSR and NMR results[18] showing that even a small amount of magnetic or non-magnetic impurity induces spin freezing at low temperatures. This behaviour was shown to scale remarkably well with that of other low-dimensional systems, independent of their geometry, such as the isolated ladder SrCu$_2$O$_3$, spin chains PbNi$_2$V$_2$O$_8$ or Spin Peierls chains CuGeO$_3$.

Thus motivated, we have carried out an extensive $^{31}$P NMR study on undoped BiCu$_2$PO$_6$ as well as BiCu$_2$PO$_6$ with non-magnetic Zn and magnetic Ni impurities doped at the Cu site to further explore such universal impurity induced features. The NMR shift and spin-lattice relaxation rate provide clear evidence for a spin-gap in the undoped compound. Our measurements clearly show an impurity-induced broadening of the spectra at low temperatures in the doped samples which is attributed to the staggered magnetisation around an impurity. This is further confirmed by Quantum Monte Carlo (QMC) simulations which demonstrate that this broadening is due to the Curie-like development of an induced spin near the impurity, whose extension saturates at low temperatures like in *S=1* spin chains[4]. Similar behavior is observed for both magnetic Ni and non-magnetic Zn impurities, implying that both impurities induce a staggered magnetization with similar extension, despite the fact that Ni $S=1$ is shown from QMC to couple ferromagnetically to its neighboring Cu $S=1/2$ spins. These results show that impurities induce the same type of (spatially) alternating magnetic moments on the sites neighbouring the impurity leading to an effectively extended moment as in uncoupled ladders or spin chains. The effective moment extension is controlled by the pure magnetic correlation length which implies a universal mechanism for a 3D spin freezing at low temperature[18].

## II. TECHNICAL DETAILS

Polycrystalline samples of $Bi(Cu_{1-x}M_x)_2PO_6$ (M = Zn, Ni) were prepared by standard solid state reaction techniques. While Zn shows complete solid solubility, Ni can be substituted up to $x = 20\%$. The details pertaining to sample preparation and characterisation are given in Ref.[19]. For the purpose of NMR, we have mainly worked on samples with $x < 0.03$. $^{31}P$ NMR measurements were done in a temperature range of *1.2-300 K*. $^{31}P$ nucleus carries a spin $I=1/2$ with *100%* natural abundance and gyromagnetic factor $\gamma/2\pi=17.254$ *MHz/Tesla*. Echo pulse NMR was used, with typical π/2 pulse widths of *3 μs*, delays between pulses of *80 μs*, and a total repetition time of *100 msec* at high temperature and up to a few seconds at low temperature when relaxation times get longer. The spectra were obtained by recombining successive Fourier Transform (FT) of the spin-echo signal obtained in a fixed field of 70 kOe by sweeping the frequency. The NMR total shift $K$ was obtained from the relative shift of the peak of the $^{31}P$ NMR line with respect to that in an $H_3PO_4$ solution used as a reference. The spin-lattice relaxation time $T_1$ was obtained using a standard saturation-recovery procedure. The recovery was found to be single exponential as expected for the $I = ½$ $^{31}P$ nucleus.

## III. PURE $BiCu_2PO_6$

The temperature dependence of the magnetic susceptibility of our undoped $BiCu_2PO_6$ sample has been reported in Ref[16]. We note that the bulk susceptibility will have different intrinsic and extrinsic contributions which are generally difficult to disentangle. In particular, analysis of low-temperature data can be troublesome when a large Curie-like term is present. In contrast, the NMR shift is insensitive to the presence of small amounts of extrinsic impurities and it measures the intrinsic spin susceptibility. The $^{31}P$ nucleus is indeed coupled to the copper spins via a super-transferred hyperfine coupling $A_{hf}$ via the oxygen (see Appendix for details). The NMR shift $K$ consists then of two parts :

$$K = K_{spin} + K_{chem} = \frac{A_{hf}}{N_A\mu_B}\chi_{spin} + K_{chem} \quad (1)$$

where $K_{chem}$ is the chemical shift, which could be measured here in $BiZn_2PO_6$. Indeed, $BiZn_2PO_6$ shows no spin contribution because of the absence of Cu, but its structure and the P orbitals are isostructural to those of $BiCu_2PO_6$. This ensures that the *T*-independent shift of only *10 ppm* measured in $BiZn_2PO_6$ is directly $K_{chem}$. This correction allows us to be confident about the "zero" of the shift (left) axis in Fig.1, where the spin shift $^{31}P$ $K_{spin}$ is compared to the measured susceptibility $\chi_{bulk}$ in undoped $BiCu_2PO_6$ as a function of temperature *T*.

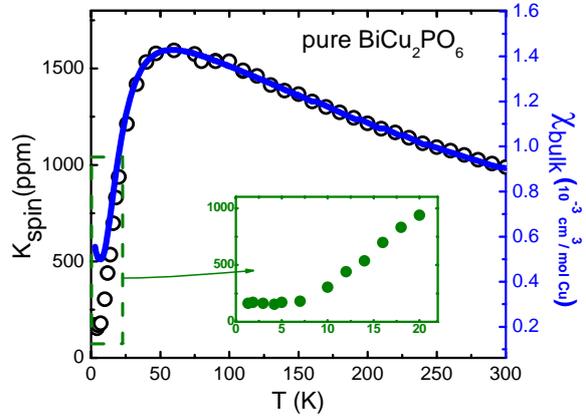

Fig.1: Variation of the $^{31}P$ NMR spin shift $K_{spin}$ (left axis) with temperature *T* is compared with that of the bulk susceptibility $\chi_{bulk}$ (right axis) for undoped $BiCu_2PO_6$. The inset zoom shows that $K_{spin}$ decreases and levels off at a small value at low-*T*.

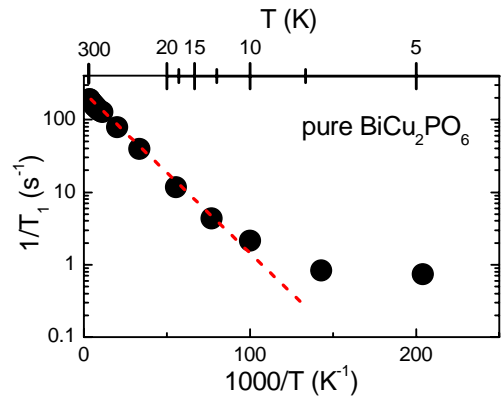

Fig. 2: The $^{31}P$ nuclear spin-lattice relaxation rate $1/T_1$ is plotted as a function of the inverse temperature on a semi-log plot, for $BiCu_2PO_6$. The solid line is a fit to activated behaviour as explained in the text. A few temperatures are shown on the top axis to provide clarity.

It is seen from Fig. 1 that $K$ scales with $\chi_{bulk}$. As was anticipated, there is a difference between the behaviour of $K$ and $\chi_{bulk}$ at low-temperature which is due to the presence of extrinsic contributions (as well as intrinsic defects/chain breaks) in $\chi_{bulk}$. On the other hand, $K$ measures the intrinsic spin susceptibility which shows no low-temperature upturn. The prominent features of the temperature dependence of $K$ are two-fold: (i) there is a broad maximum (at about 60 K) which is a hallmark of low-dimensional magnetic systems, and (ii) there is a sharp decrease below the maximum, with $K$ becoming small as $T$ approaches zero, due to a gap in the spin excitations. Fitting the shift data at low temperatures to $K(T) = K_0 + A\exp(-\Delta/k_B T)/\sqrt{T}$ which is valid for one-dimensional spin-gapped systems,[20] yields $\Delta = 37 \pm 5 K$ for the gap $\Delta/k_B$. The reason for a non-zero residual value for $K_0$ (~ 200 ppm) is not clear at present but is not due to any chemical shift contribution as explained hereabove.

The nuclear spin-lattice relaxation rate $1/T_1$, which probes the low-energy spin dynamics in a system, has been known to provide important information in a variety of low-dimensional spin systems such as the spin chain $Sr_2CuO_3$ [21], the two-leg ladder $SrCu_2O_3$ [22], the Haldane chain $Y_2BaNiO_5$ [23], etc. Our $^{31}P$ $1/T_1$ data (see Fig. 2) show an activated $\exp(-\Delta_{1/T1}/k_B T)$ behaviour as expected for a gapped system, and we find that $\Delta_{1/T_1} = 50 \pm 5 K$. The $T$-dependence of $1/T_1$ in one-dimensional spin-gapped systems[20] is indeed expected to follow $1/T_1 \propto \sqrt{T}\exp(-3\Delta/2k_B T)$. Therefore, in the $k_B T/\Delta \ll 1$ limit, the leading behaviour of $\chi$ and $1/T_1$ is going to be $\exp(-\Delta/k_B T)$ and $\exp(-\Delta_{T1}/k_B T)$ respectively, with $\Delta_{T1}/\Delta = 1.5$. This ratio is consistent with our findings within error bars. The deviation of $1/T_1$ from the exponential decrease at low-$T$ might be due to intrinsic defects.

So, both the static spin susceptibility and the dynamic relaxation rate NMR measurements consistently demonstrate that the system is spin-gapped, as expected in such a ladder geometry.[6]

**IV. NMR STUDY OF THE IMPURITY EFFECTS**

$^{31}P$ NMR spectra were measured for Ni and Zn substituted $BiCu_2PO_6$ at various temperatures. Typical evolution of the spectra is reported in Fig. 3. When temperature is lowered, the spectrum shifts toward lower frequencies and strongly broadens. When a local defect is created at a Cu site, NMR enables one to measure the histogram of the local fields both close to and far from the defect through $^{31}P$ NMR hyperfine coupling to all the Cu sites. As demonstrated in various low-dimensional AF chains, ladders, or planes, the center of gravity of the spectra, i.e. their shift, is proportional to the spin susceptibility *far from the defects*[1]. In contrast, any induced magnetism close to the impurity is reflected, either in a broadening of the spectrum or in the appearance of satellites in the wings of the spectrum. The former is exactly the behavior observed in Fig.3.

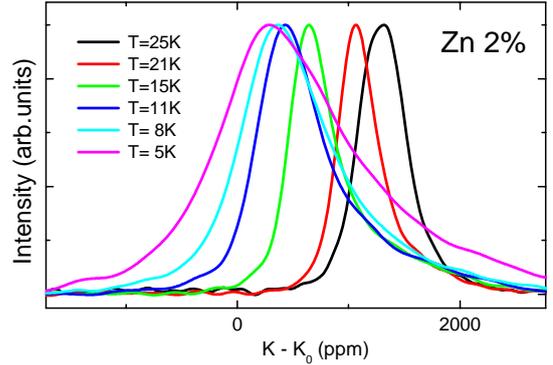

Fig. 3 : $^{31}P$ NMR spectra for 2% Zn doped $BiCu_2PO_6$ at various temperatures. The spectrum moves to smaller shifts and broadens as temperature is lowered.

A. Susceptibility far from the defects

The $^{31}P$ shift plotted in Fig. 4 is found almost identical for substituted and pure materials. This is direct evidence that the uniform spin susceptibility $\chi_0$ far from the defects is unaffected either by Zn or Ni. So the ladder spin-gap estimated to be *35-40 K* in the pure compound remains identical in the presence of impurities. A very slight increase of the low-temperature shift is observed for Zn:*2%* (see the inset of Fig.4). This could be due to the presence of some in-gap, impurity-induced density of states but, given the experimental accuracy, the effect is too small to allow firm conclusions.

A comparison of the $^{31}P$ NMR shift with macroscopic susceptibility can be made, using the hyperfine coupling estimated in pure $BiCu_2PO_6$, as reported in the inset of Fig.4. At temperatures lower than that of the broad maximum, down to *T=25 K*, the macroscopic susceptibility and the $^{31}P$ NMR shift both follow a similar decrease when decreasing temperature, due to the opening of the spin -gap.

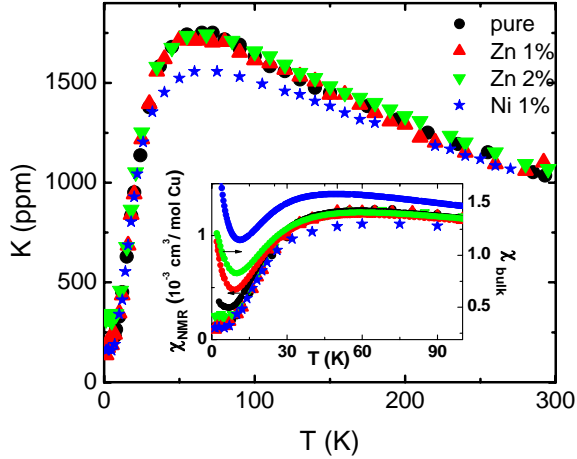

Fig.4 : The $^{31}$P NMR shift of BiCu$_2$PO$_6$ as a function of $T$ is not seen to be significantly changed by Zn or Ni doping while the bulk susceptibility (right axis in the inset) has a marked low-$T$ increase.

### B. Induced staggered magnetization close to the defects

At temperatures lower than $T=25K$, the NMR spin susceptibility far from the defects saturates to almost zero while the macroscopic susceptibility diverges, typically as $1/(T+\theta)$. In the same low-temperature regime, the NMR linewidths increase with a similar $T$-behavior, as shown in Fig.5. This low-$T$ Curie-like increase in macroscopic susceptibility and the NMR broadenings signal the appearance of induced, staggered, paramagnetic moments near each impurity. Both effects scale with the impurity content as expected. Similar features were observed in various compounds with non-magnetic or magnetic impurities, such as spin chains,[2] spin ladders,[5] and 2D planes (high-$T_c$ cuprates).[1,24]

For a quantitative comparison, we first have to correct the NMR linewidths for other sources of broadening. The inset of Fig. 5 shows that the linewidth and the shift scale with each other above $T = 30\ K$, when the impurity effect is not too large. It means that the NMR line has an intrinsic, small width proportional to the shift. It is likely due to the random orientation of the crystallites composing our samples, and which in turn gives rise to a distribution of the shift itself. We used this scaling to extrapolate this "intrinsic" contribution to the width down to $T = 0$ and subtracted it from the full linewidth. The resulting impurity-only contribution to the width $\Delta\nu_{impurity}$ is plotted on Fig.6. As seen, it is sizeable only at low-temperatures and scales well with the impurity content, as expected. Note that a Curie behavior is also observed for the width of the pure compound, which implies that there are some intrinsic defects in the material such as vacancies or other types of local disorders, as is usual in these oxides[2]. This allows us to estimate the typical amount of such native defects to be about $x=0.1$ to $0.3\%$, a value which agrees well with macroscopic measurements reported on the same samples.[19]

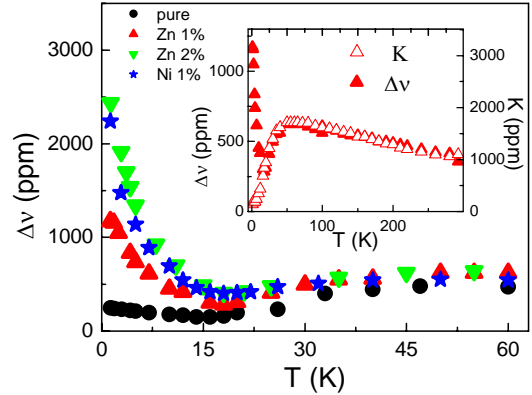

Fig. 5: The measured full width at half maximum (FWHM) of the $^{31}$P NMR spectrum is plotted as a function of $T$ for undoped, Zn-doped (1 and 2 %), and Ni-doped (1 %) BiCu$_2$PO$_6$. As seen in the inset, the FWHM scales with the shift, except in the low-$T$ region. This has been used to extract the dopant-induced FWHM in the various samples, as explained in the text.

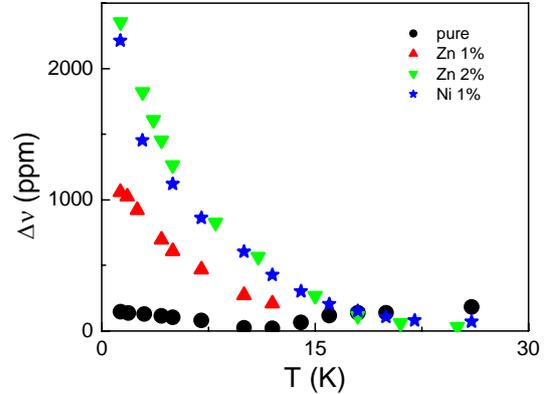

Fig. 6: The $T$-dependence of the impurity-only contribution to the width $\Delta\nu_{impurity}$ is plotted for different samples. The "intrinsic" width was subtracted from the measured FWHM as explained in the text.

## V. QUANTUM MONTE CARLO VERSUS NMR

QMC is a very precise tool to investigate localized moments physics in spin gapped materials like spin ladders [25], Haldane chains [26,4], or coupled ladders systems[27]. Here, we employ Stochastic Series Expansions (SSE) method[28,29] to get a better understanding of how staggered moments develop in a coupled ladders model perturbed by a finite concentration of spinless (Zn) or $S=1$ (Ni) impurities and how this modifies NMR spectra.

### A. The model

The starting point is a realistic model of AF coupled spin-1/2 ladders that we previously investigated in Ref.[18]:

$$H = \sum_{\langle ij \rangle} J_{leg} \vec{S}_{i,j}.\vec{S}_{i+1,j} + J_{rung} \vec{S}_{i,2j}.\vec{S}_{i,2j+1} \\ + J_\perp \vec{S}_{i,2j+1}.\vec{S}_{i,2j+2} - g\mu_B H \sum_i S_i^Z \quad (2)$$

This model[30,31] displays a gapped valence bond solid (VBS) phase when the inter-ladder coupling $J_\perp$ is not too strong: for isotropic ladders ($J_{rung} = J_{leg}$), VBS is achieved whenever $J_\perp < J_\perp^c$ with $J_\perp / J \approx 0.314$ [30]. Previous bulk studies [16] on BiCu$_2$PO$_6$ estimated $J = J_{rung} = J_{leg} \approx 100K$. Muffin-tin orbital calculations suggest the existence of an additional frustrating coupling along the ladders [16]. We did not take this possible frustration into account in our model Eq. (2) as we focus on capturing semi-quantitative features of the low-T physics i.e. a weakly coupled ladders system having a sizable spin gap. The system of coupled ladders with static spinless impurities is depicted in Fig.7. In BiCu$_2$PO$_6$, the experimental spin gap $\Delta \approx 35 - 40K$ yields an inter-ladder coupling $0.1 \leq J_\perp / J \leq 0.2$ for the theoretical model under study Eq. (2)[18].

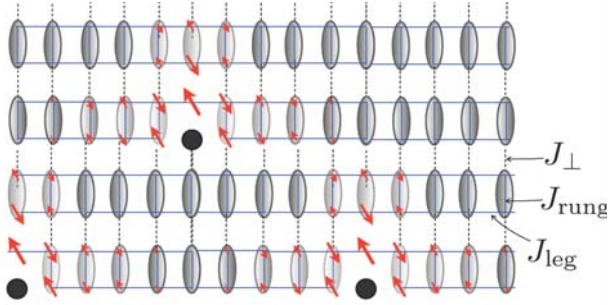

FIG. 7: (Color online). Schematic picture for the model Eq. (2) of coupled ladders having three different couplings $J_{leg}, J_{rung}, J_\perp$. Focusing in the gapped regime ($J_{leg} = J_{rung} = 10J_\perp$), the system displays a VBS ground-state illustrated by rung singlets. Each non-magnetic impurity (black dots) breaks such a rung singlet and releases a spin 1/2 degree of freedom that is exponentially localized around the vacancy with an AF pattern.

Therefore this model is a good starting point to investigate the effects of a finite concentration $x$ of non-magnetic impurities on a gapped system of coupled ladders, and try to obtain NMR spectra from QMC calculations. We chose the following parameters: $J = 10J_\perp = 100K$ which implies a spin gap $\Delta \approx 37K$.

### B. Simulated NMR spectra for Zn doping

To get NMR spectra from QMC simulations, we first impose a finite magnetic field $H = J_{leg}/10$, corresponding to the typical field used in the NMR experiments of about $7\,T$. Using the standard directed loop SSE algorithm[28], we compute the local magnetizations $\langle S_i^Z \rangle$ for samples of size $L \times L$ with $x = 2\%$ of vacancies. Simulations are performed at various temperatures $T = 5, 10, 20, 40, 60, 80, 100\,K$ over an ensemble of $N_s$ independent disordered samples in order to get a good statistic over the random doping process. As we are interested in the distribution of local magnetizations (which will map onto the frequency-dependent NMR spectrum as explained below), the number of Monte Carlo steps $N_{MC}$ devoted to the measurement turns out to be a crucial quantity since a high precision is required on each of the local observables $\langle S_i^Z \rangle$. We have checked that the statistical errors decay like $1/\sqrt{N_{MC}/L^2}$ (which is the square root of the inverse number of MC steps per site). Therefore, we performed $1.6 \times 10^7$ measurement steps for square lattices of size $32 \times 32$ in order to reduce the contribution from statistical fluctuations in the distribution of local fields. Disorder averaging has been performed over $N_s=200$ independent randomly doped samples. At a given temperature $T$, local magnetizations $\langle S_i^Z \rangle$ are computed for each randomly doped sample. Once data from a sufficiently large enough number of independent samples are obtained, an histogram of the Knight Shifts $P(K_i)$ can be drawn (see Fig. 8) with no additional free parameter, using Eq.(5) which relates the $^{31}$P NMR shift and the local magnetizations.

While quite simple, the model Eq. (2) captures qualitative and even semi-quantitative features of experimental spectra where one can observe (Fig. 9) shift and broadening temperature dependences similar to the experimental ones.

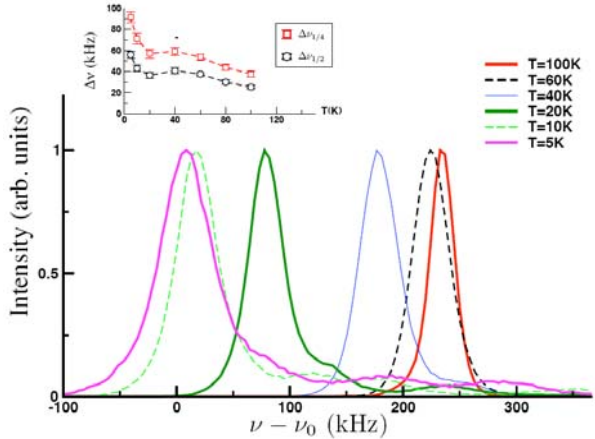

FIG. 8: (Color online). Main panel: theoretical NMR spectra from QMC simulations performed for a simple model of weakly coupled S=1/2 ladders [Eq. (2)] with x = 2% of non-magnetic (Zn-like) impurities. Inset: full widths at half $\Delta\nu_{1/2}$ and quarter $\Delta\nu_{1/4}$ maximum versus T, rapidly growing below the spin gap $\Delta \approx 37K$ in qualitative agreement with experimental data.

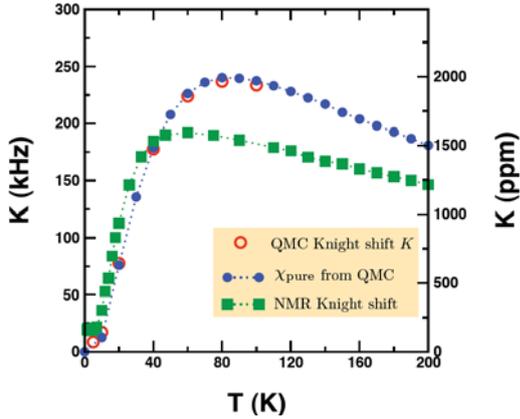

FIG. 9: (Color online) Temperature dependence of the uniform susceptibility computed using QMC for a 32 × 32 sites system of weakly coupled ladders [model Eq.(2)].)] with $J = 10 J_\perp = 100 K$. Blue dotted line shows the results of $\chi(T)$ obtained on the pure system while red stars show the knight shift computed for doped samples (x = 2% of S = 0 Zn impurities). Experimental NMR Knight shifts are also shown for comparison (green squares).

As a first theoretical check, we have compared the T-dependence of the average shift position $K(T)$ (which represents the magnetic response far from the defects) with the spin susceptibility of the pure system $\chi(T)$ computed independently with QMC on a pure system. These results are shown in Fig. 9 where, as expected, a good agreement is observed between theoretical $K$ and $\chi$. Both $K$ and $\chi$ are also consistent in magnitude and behavior with the experimental NMR shifts (green squares) but show deviation from the experiment at intermediate temperature around the maximum and at higher temperatures. Such a discrepancy is attributed to frustration effects, not included in our theoretical model, which are known to reduce the maximum of $\chi$ [32].

### C. Impurity effects from QMC simulations

The broadening of the local fields distribution is a direct signature of the impurity-induced paramagnetic effects in a magnetic field when the system is cooled down below the spin gap $\Delta$. This broadening is observed in the QMC spectra at low $T$ as seen in the inset of Fig. 8 where the full widths at half $\Delta\nu_{1/2}$ and quarter $\Delta\nu_{1/4}$ maximum rapidly grow at low $T$, in qualitative agreement with experiments (fig.6). However, our simple 2D model does not allow for a direct quantitative comparison between QMC and NMR since frustration and, more importantly, finite-$T_g$ 3D ordering are not included in the model. Beyond these discrepancies, we can discuss the general broadening mechanisms at play in the real 3D material.

For $T \leq \Delta$, exponentially localized induced moments start to build up in the vicinity of each dopant. For a 3D system the average moment profile displays the 3D localized form

$$\langle S^z(\vec{r}) \rangle = (-1)^{x+y+z} S_0 \exp(-\frac{x}{\xi_x} - \frac{y}{\xi_y} - \frac{z}{\xi_z}) \quad (3)$$

Here $\xi_x, \xi_y, \xi_z$ are the correlation lengths for AF order in the three spatial directions. They are in general all different due to the anisotropy of interactions. The effect of the impurity is to induce an effective moment of typical size $S_0$ with a 3D spatial extension $V_\xi \propto \xi_x \xi_y \xi_z$ around the impurity. In the dilute limit ($V_\xi \ll 1/x$), we expect the width of the local fields distribution to scale with the impurity concentration $x$, the typical size of the induced moment $S_0$ and its 3D spatial extension $V_\xi$:

$$\Delta\nu \propto x S_0 V_\xi(T). \quad (4)$$

This relation can be deduced intuitively in the following way. The broadening is due to unresolved satellites corresponding to the induced moments on spins close to the impurity. Their intensity scales with $x$ and their number scales with $V_\xi$. These are not resolved here because of the dimensionnality and hyperfine factor of $^{31}P$ which averages over 4 sites.

Two main sources of broadening are therefore showing up upon cooling of the system: a rapid growth of (i) the extension and (ii) the size of induced moments.
(i) Regarding the spatial extension $V_\xi(T)$, while a rapid increase is expected below $T \sim \Delta$, a transient saturation regime appears before 3D effects at lower T come into play, formally leading to a diverging $V_\xi(T)$ when $T \rightarrow T_g$. Note however that strongly anisotropic effects are also expected, leading to a more complicated T -dependence of $V_\xi(T)$.

(ii) Concerning the typical size $S_0$ of the induced moments, below the spin gap there is a free impurity spins regime [8] where the external field induces a Brillouin response which leads to a Curie-like behavior $S_0 \propto H/T$, in the range $\Delta > T >> H$, and saturates at lower $T$. Note however that frustration combined with random dilution may also lead to a reduction of the effective moment at lower temperature [33].

Taken all together, these various effects imply that the exact temperature (and field) broadening of the NMR lines is a quite subtle issue that will be investigated further in a future work.

### D. Comparison between Zn and Ni

We now turn to the difference between non- magnetic Zn and magnetic $S=1$ Ni-doped samples. As shown in Fig.6, there is roughly a factor *2* between experimental NMR linewidths ($\Delta\nu_{Ni}/\Delta\nu_{Zn} = 1.9$). However, the spatial extension of induced moments does not depend on the magnetic nature of the impurities, as shown in Fig.10. Indeed, such an extension is basically controlled at low enough temperature by the inverse bare spin gap $1/\Delta$. Therefore this difference has to come from the moment size itself $S_0$. Such a conclusion is also reached by the analysis of Curie tails performed in Ref.[19] where a reduced total moment of $S_{Zn}^{tot} \approx 0.35$ and $S_{Ni}^{tot} \approx 0.79 \approx 2.2 S_{Zn}^{tot}$ is measured by macroscopic magnetization.

In order to understand the physical origin of such an effect, we performed QMC simulations on a simple model of isolated ladders and compared magnetization profiles for $S_{imp} = 0$ or $S_{imp} = 1$ coupled to the bulk spins-1/2 with a coupling $J'$ that may be different from $J$ (depicted on top of Fig. 10). Results of such a computation are shown in Fig. 10 for three representative cases: $S_{imp} = 0$, $S_{imp} = 1$ for an antiferromagnetic AF $J'>0$ and a ferromagnetic F $J'<0$ coupling with Cu. As stated above, the induced moment extensions are roughly identical: $\xi_{Ni} \approx \xi_{Zn}$. However, the prefactor $S_0$ of the exponential decay does depend on the nature of the coupling $J'$ between Cu ($S = 1/2$) and Ni ($S = 1$).

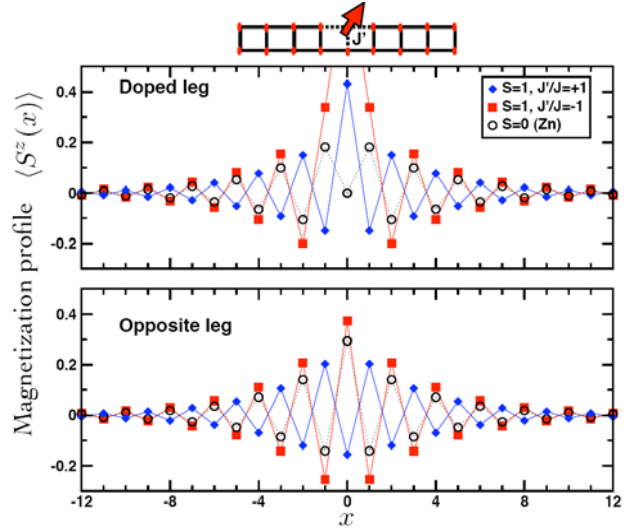

FIG. 10: (Color online) Magnetization profiles of an isolated spin-1/2 ladder with a single spin-S impurity (depicted on the top) coupled with J' to the rest. QMC results obtained on 32 × 2 samples at T = J/100 and H = J/10. Three cases are shown here: (i) S = 0 (open circle, Zn case) and (ii) $S_{imp}$ = 1 with J'=J (blue diamond) lead to quantitatively similar features, while the third case (iii) $S_{imp}$ = 1 and J'=−J (red squares) shows an enhanced effect.

The cases $S_{imp} = 0$ and $S_{imp} = 1$ for an antiferromagnetic AF $J'$ display similar behaviors (except a trivial parity effect: see Fig. 11) since the ground-state lies in the $S_z=1/2$ sub-sector. On the contrary, when $S_{imp} = 1$ is ferromagnetically coupled to its neighboring spins 1/2, the ground-state is in the sub-sector $S_z=3/2$ which leads to an enhancement of the magnetization profile, without changing the spatial extension $\xi$. To go further, we have performed systematic QMC computations for the magnetization profiles on isolated ladders for $S_{imp} = 1$ and various $J'$ and compared the results to the $S_{imp} = 0$ case. After performing an exponential fit, we get the amplitudes $S_0^{Ni}$ plotted in Fig. 11 versus $J'$. The experimentally observed factor *2* is obtained here for a ferromagnetic coupling $J'$ of the order of $J$ at least. Note that simulations are performed on isolated ladders. In the real compound, we expect dimensionnality and frustration effects to enhance correlations and therefore $S_0$, leading to a smaller absolute value of $J'$ to account for the experimental findings. While it is hard to check these findings from simple quantum chemistry arguments, we stress that frustration combined with a ferromagnetic coupling between Ni and Cu will naturally enhance the effective moment amplitude around a Ni impurity.

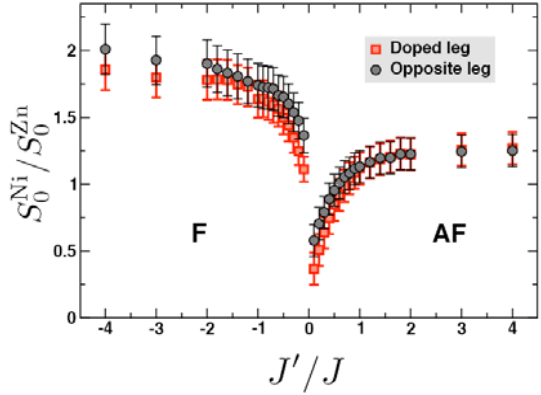

Fig. 11: (Color online) QMC results for the amplitudes ratio $S_0^{Ni}/S_0^{Zn}$ of the magnetization profiles (doped and opposite legs) versus the coupling J' between Ni and Cu.

Indeed, as schematized in fig.12, frustration which exists in pure BCPO because of a second neighbor AF coupling $J_4$ between Cu would be removed here locally near Ni. Such an effect will naturally enhance local moment amplitudes.

Therefore, while it is hard to get a precise estimate of how Ni couples to the neighboring Cu in Ni-doped BiCu$_2$PO$_6$, the QMC simulations and NMR experiment suggest a substantial ferromagnetic coupling $J'$ of the order of $J$.

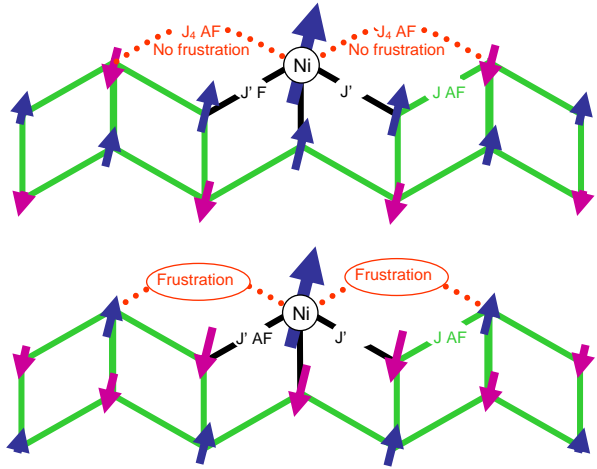

Fig.12: (Color online) Ni effect on the spin ladder in case of a ferromagnetic (upper panel) or antiferromagnetic (lower panel) coupling with neighbor Cu. For a $J_4$ AF next neighbor coupling, frustration is removed by Ni for the ferromagnetic case.

## VI. DISCUSSION

We have shown that in the coupled spin ladder BiCu$_2$PO$_6$, non magnetic and magnetic impurities both induce staggered moments which develop below the spin gap. Their extension is similar, only their amplitude differs, because Ni couples ferromagnetically to its neighboring Cu, while Zn does not couple at all. These results can be compared to similar studies in other low dimensional quantum magnets.

In isolated spin ladders SrCu$_2$O$_3$, Cu NMR studies reveal similar qualitative behavior: both Zn and Ni induce Curie-like paramagnetic NMR broadenings which signal staggered moments[5,6]. But opposite to our case, Ni is found to induce a broadening identical to Zn[6]. In these isolated ladders, Ni is indeed probably coupled to its adjacent Cu with an AF coupling similar to that of the pure ladder, leading to same effect than Zn. This contrasts with BiCu$_2$PO$_6$ where the likely ferromagnetic effective Ni-Cu couling, in concomitance to large inter ladders couplings and possibly frustration effects, produces an induced moment larger for Ni than for Zn. While qualitative results are similar in our study and Ref.[5], interpretations are very different. In Ref.[5], it has been argued that the moment extension $\xi$ depends on impurity content as $\xi \sim 1/x$. This would contradict our own interpretation and QMC results where $\xi$ is found to be independent of impurity content, and governed solely by the intrinsic correlation of the pure ladder. But this $\xi \sim 1/x$ conclusion comes mostly from the data taken at a very small Zn content $x=0.1\%$ where broadening is not scaling with impurity content anymore. However, at such small content, the pure linewidth and the impurity effect are so small that incertitude is too large to determine the impurity-induced broadening safely. All other data of Ref.[5], and all data reported in an independent paper[6] on same compounds are consistent with our own interpretation of $\xi$ independent of impurity content and intrinsic to the ladder's physics.

In spin $S=1$ Ni Haldane chains, the very same trends are observed: both non magnetic Zn and magnetic $S=1/2$ Cu at Ni site induce similar exponentially decaying staggered moments, with same extension $\xi$ and slightly different amplitudes[4], $\xi$ being hereagain identical to the pure correlation length.

Finally, underdoped high Tc cuprates such as YBa$_2$Cu$_3$O$_{6+y}$ (0.4<y<0.9) can be considered as a realization of a strongly antiferromagnetically correlated two dimensional Cu system, even though they are also metallic. Here again, both non magnetic Zn or magnetic Ni substituted at Cu site of CuO$_2$ two-dimensional $S=1/2$ layers induce similar staggered moments with the same temperature dependencies but different amplitudes[1,24].

Our results suggest a possible universal picture which could apply to quantum antiferromagnetically coupled low dimensional systems, the BiCu$_2$PO$_6$ bridging the 1D physics of spin chains and isolated ladders to the 2D physics of cuprates. Any non magnetic as well as magnetic impurity induces a staggered moment at temperatures smaller than the relevant energy scale (magnetic coupling and/or spin gap). This moment follows a Curie-like dependence which may be affected if the extension ξ also depends on temperature. Only the amplitude differs in the case of a magnetic impurity, depending on the size and sign of the coupling between this impurity and neighboring spins. The extension ξ is characteristic of the intrinsic system and does not depend on impurity content at small impurity content. At low enough temperature, because of the other residual 3D couplings present in the compound, these induced extended moments may interact enough to freeze in an antiferromagnet or a cluster-type antiferromagnet, in a universal fashion as well. This generic scenario needs to be further confirmed in other low dimensional systems with different geometries and couplings.

Precise understanding of how this generic 3D ordering takes place remains an open issue related to the important question of criticality close to a transition, and needs more theoretical and experimental effort.

## ACKNOWLEDGMENTS

We acknowledge H. Alloul, M. Azuma, S. Capponi, I. Dasgupta, , B. Grenier, Y. Kitaoka, T. Masuda, P. Mendels, D. Poilblanc, V. Simonet, A. Zorko for very helpful discussions. This work was supported by the EC FP 6 programme, Contract No. RII3-CT-2003-505925, ANR Grant No. NT05-4-41913 ''OxyFonda,'' the Indo-French center for the Promotion of Advanced Research, and ARCUS Ile de France-Inde. Some of the calculations were performed using the SSE code[29] of the ALPS libraries [34].

## APPENDIX : HYPERFINE COUPLINGS IN BiCu$_2$PO$_6$

In this appendix, we demonstrate from NMR experiments that each P nucleus is coupled through hyperfine coupling to two Cu atoms of the adjacent ladder as shown in fig.14. The *total* hyperfine coupling $A_{hf}$ between one $^{31}$P nucleus and the neighbouring Cu, defined by $K_{spin} = A_{hf} \chi_{Cu} / \mu_B$ (where $\chi_{Cu}$ is the atomic susceptibility pertaining to each Cu), is determined experimentally from Fig.1 to be $A_{hf}^{tot} = 0.652\, T$. The spin density largely resides on the Cu orbitals and an effect (in terms of an NMR shift) is felt at the $^{31}$P site due to a hybridisation of the P 4$s$ orbitals with the Cu $d$ orbitals via the oxygen. We want to determine which hybridization actually contributes to this coupling. The various possible hybridization paths between $^{31}$P and adjacent Cu are plotted on Fig.13. The band structure calculations reveal that in the global (crystallographic) co-ordinate system the Cu $yz$ orbitals lie near the Fermi level, so we focus on this orbital only and its effective hybridization with P−$s$ orbitals[35]. Because of symmetry reasons, this hybridization along path labelled α and β is zero. Therefore, *each P is coupled to four Cu all from the same ladder* through the path labelled *a*, as schematically displayed in Fig. 13.

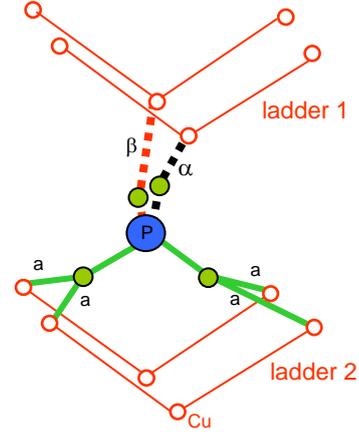

Fig. 13: the dominant hybridization paths between one $^{31}$P (in blue) and its adjacent Cu (in red) through oxygen orbitals (in green) are represented. For symmetry reasons, there is no hybridisation via paths α and β.

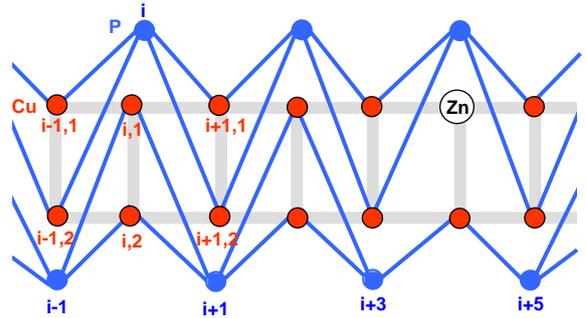

Fig. 14: Hyperfine couplings between P (blue) and Cu (red) are symbolized by the blue lines. In the presence of a Zn, the P labelled *i+3* and *i+5* probe only 3 Cu.

This results for each P nucleus at site $i$ in a shift:

$$K_{spin}^{i} = A_{hf}^{per\,Cu}(\chi_{i-1,1}^{Cu} + \chi_{i+1,1}^{Cu} + \chi_{i-1,2}^{Cu} + \chi_{i+1,2}^{Cu})/\mu_B \quad (5)$$

where $A_{hf}^{per\,Cu} = 0.163 T$ is the hyperfine coupling between one $^{31}$P and one Cu via path $a$. This can be checked experimentally from the analysis of Zn substitution effect on the $^{31}$P NMR spectrum. When the temperature is much larger than the gap opening, no induced moment is expected from Zn substitution at the Cu site. For one Zn defect, the only effect is the removal of one Cu and of its contribution to the near neighbour (nn) $^{31}$P shift coupled to this Zn (Fig.14). If each $^{31}$P is coupled to N Cu, the P nn to Zn should lead to a distinct less shifted satellite line, with a shift

$$K_{nn} = \frac{N-1}{N} K_{main}$$

The relative intensity of this line is then directly linked to the concentration $x$ of Zn per Cu:

$$I_{nn} = Nx(1-x)^{N-1} I_{whole\,spectrum}$$

In the rare case where two Zn are adjacent, an even smaller and less shifted $^{31}$P satellite line (noted with an asterisk) should result with:

$$K_{nn}^* = \frac{N-2}{N} K_{main} \quad and \quad I_{nn}^* = \frac{N(N-1)}{2} x^2 (1-x)^{N-2} I_{whole\,spectrum}$$

The $^{31}$P NMR spectrum at $T=300K$ plotted in Fig.15 indeed displays these two reduced and less shifted satellites in the presence of Zn. The measured satellite shifts $K_{nn}$ and $K_{nn}^*$ are found reduced compared to $K_{main}$ by a ratio *0.7* and *0.4* respectively which points to N between *3* and *4*. But Zn could slightly distort its environment and change the local hyperfine couplings, which would in turn affect these ratios. On the contrary, the intensities of these satellites plotted on fig.16 depend only on the Zn concentration $x$ which makes it a more reliable quantity. A good agreement with *N=4* is found for both satellites, with no adjustable parameter. This unambiguously confirms that the picture depicted in Fig.14 is correct, where each P is coupled to the 4 Cu of a single ladder. The hybridization ε between Cu d and P 3s orbitals at play in the hyperfine coupling path $a$ can be roughly estimated using:

$$A_{hf}^{per\,Cu} \approx \varepsilon^2 A_{hf}^{3s}$$

where $A_{hf}^{3s} \approx 210\,T$ [36]. This leads to $\alpha \approx 3\%$, which is reasonable in view of ab-initio estimates[35].

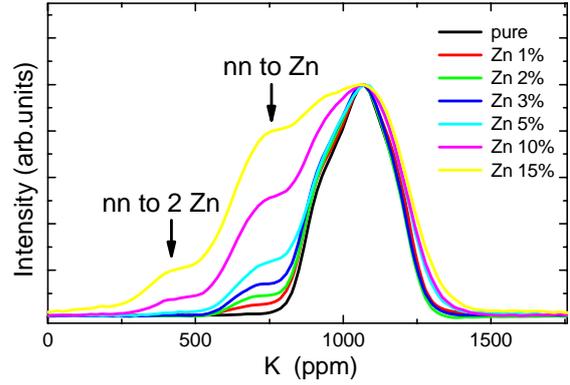

Fig. 15: The $^{31}$P NMR lineshapes for Zn doped BiCu$_2$PO$_6$ samples at 300 K. The shoulders are identified as arising from $^{31}$P nuclei near the Zn as explained in the text

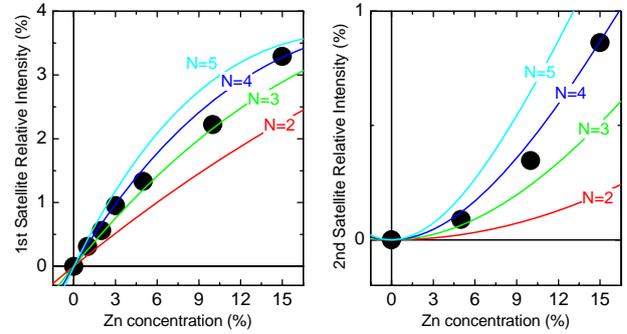

Fig. 16: The relative intensities of the two satellites shown in Fig.11 for various Zn contents, measured at $T=300K$.

*Corresponding author : bobroff@lps.u-psud.fr